# Collective Dynamics in Spiking Neural Networks: A Systematic Review

Afifurrahman,[1, a)] Mohd Hafiz Mohd[2, b)]

[1)] *Tadris Matematika, Universitas Islam Negeri Mataram, Mataram 83116, Indonesia*
[2)] *School of Mathematical Sciences, Universiti Sains Malaysia, USM 11800, Penang, Malaysia*

[a)] *Corresponding author: afif.rahman@uinmataram.ac.id*
[b)] *mohdhafizmohd@usm.my*

**Abstract.** This study aims to identify typical collective phenomena that emerge in excitatory and inhibitory (E-I) spiking neural networks as reported in recent computational studies. The research methodology used is Preferred Reporting Items for Systematic Reviews and Meta-Analyses (PRISMA) procedures, comprising three primary stages: an initial search for literature in the SCOPUS database, a screening process based on specific inclusion and exclusion criteria, and a review of the selected literatures. Out of 491 documents from 2014 to 2024, six research papers are selected in the review stage. Four generic collective regimes have been identified, including synchrony, irregular behavior, stationary state, and oscillatory patterns. Our review findings suggest that the collective dynamics of E-I spiking neurons stem from the interplay of intrinsic neuronal characteristics, network topology, and external stimuli. Additionally, the prevalent use of Quadratic Integrate-and-Fire (QIF) neuron model in the literature highlights its significance as a robust candidate for exploring collective behaviors in large-scale neuronal networks. The findings outlined in this paper might be useful for individuals who lack prior familiarity with computational modelling of spiking neurons but have an interest in the field.

## INTRODUCTION

### Background

A neuron is a fundamental processing unit within the central nervous system. The dynamics of neurons are characterized by changes in the membrane potential (voltage). When the membrane potential surpasses a certain threshold, the neuron fires a spike (or pulse), after which it repolarizes in preparation for the next spike. The interval between consecutive spikes is known as the inter-spike interval. In the absence of coupling, neurons fire at a constant rate if they operate above the threshold. Neurons communicate through these pulses and are categorized into excitatory and inhibitory types. Excitatory neurons produce positive spikes that increase the firing rate of the receiving neurons, while inhibitory neurons generate negative spikes that reduce firing activity.

Neurons are embedded in networks of billions of neurons organized into different brain regions with specific type and role [1]. From a dynamical perspective, the key question focuses on identifying the collective behaviours that emerge spontaneously and their potential links to various brain functions. To address such question, spiking neural networks (SNNs) are often used in theoretical and computational studies. SNNs are a class of artificial neural networks that aim to mimic the way biological neurons communicate through discrete spikes or action potentials [2,3]. Unlike traditional artificial neural networks, which operate on continuous values and rely on activation functions, SNNs process information through the timing and frequency of spikes, making them closer to the biological processes observed in the brain.

In this paper, we conduct a systematic review on the literatures discussing the spontaneous emergence of collective phenomena in spiking neural networks (SNNs). The term *collective* refers to substantial changes on the systems state induced by mutual coupling [4]. To the best of our knowledge, no specific research has been undertaken to provide a *systematic* review of the emergence of collective dynamics in spiking neural networks. Therefore, it is essential to undertake this research to deliver a thorough review of the associated research theme and promote the up-to-date results to broader audience of computational neuroscience. Our methodology adheres to the Preferred Reporting Items for Systematic Reviews and Meta-Analyses (PRISMA) guidelines [5]. We concentrate on studies that explore collective behaviors in two interacting populations of excitatory and inhibitory neurons, where individual neurons are modeled using spiking neuron models discussed in [6]. We aim to address the following questions: (i) What types of dynamical regimes arise in excitatory-inhibitory (E-I) neural networks? (ii) What are the minimal conditions required for their existence? The literature search encompasses unit cells in the E-I network, specifically involving integrate-and-fire (IF) and Hodgkin-Huxley (HH) neurons. These two foundational neuron models are commonly employed in simulations and are highly effective in capturing emerging dynamical behaviours [6,7]. These models are discussed briefly in the next subsection.

## Models

### *Integrate-and-Fire Neuron*

Since Lapique's work in 1907 [8], integrate-and-fire models have still gained interest among researchers. The simplest model in class of IF neuron is called leaky integrate-and-fire (LIF) model consists of two main components: (i) linear differential equation describing the evolution of membrane potential $u(t)$ and (ii) a threshold for spike firing. If $u(t)$ reaches a threshold $\vartheta$, the neuron is said to fire a spike at $t_s$ and $u(t)$ is instantaneously reset to $u_r$ at $t = t_{s_+}$. The equation is given by one-dimensional system [9-11]

$$\dot{u} = -u(t) + I(t) \tag{1}$$

combined with a reset condition: if $u(t_s) \geq \vartheta \Rightarrow u(t) = u_r$ for $t = t_{s_+}$. Here, $\dot{u}$ is a derivative of $u$ with respect to time $t$. In the absence of input current $I(t)$, $u(t)$ decays to the resting membrane potential as $t \longrightarrow \infty$.

A particular generalization of IF neuron is quadratic integrate-and fire (QIF) model, given by nonlinear differential equation [9-11]

$$\dot{u} = u^2(t) + I(t) \tag{2}$$

while if $u(t_s) \geq u_{peak} \Rightarrow u(t) = u_r$ assuming that $u_r = -u_{peak} = -\infty$. When $u(t)$ goes to infinity, we say that the spike has been fired. In numerical simulation, however, the peak value $u_{peak}$ must be finite, so it is convenient to normalize it to $u_{peak} = 1$ [11].

*Hodgkin-Huxley Neuron*

In 1952 [12], Alan Hodgkin and Andrew Huxley proposed a mathematical model describing initiation and propagation of action potential (spike) via ion channels in an isolated neuron of squid giant axon. It is defined as a four-dimensional nonlinear system [9-11]

$$\begin{cases} C\dot{u} = -g_L[u(t) - E_L] - g_{N_a} m^3 h\left[u(t) - E_{N_a}\right] - g_K n^4[u(t) - E_K] + I(t) \\ \dot{x} = \alpha_x(u)[1 - x] - \beta_x(u)x, \quad \text{for } x = m, n, h. \end{cases} \tag{3}$$

$C$ is a capacitor and parameters $g_L, g_{N_a}, g_k$ are the conductance for leak, sodium, and potassium channel, respectively. The parameters $E_L, E_{N_a}, E_k$ represent Nernst equilibrium potential for leak, sodium, and potassium channel, respectively. The variables $m(t), n(t), h(t)$ control the opening (activation) and closing (inactivation) of ion channels, while the functions $\alpha_x(u)$ and $\beta_x(u)$ characterize the voltage-dependent transition rates [9].

Fast and slow oscillations are prominent features in the Hodgkin-Huxley model, if a constant input current $I(0) = I_0$ exceeds some critical value. Based on this observation, the model in principle can be reduced to a two-dimensional system considering only fast and slow variables (see [13]).

### *Network Structure*

The models (1)-(3) describe the time evolution for a single neuron. While the brain contains billions of neurons organized into various regions, it is important to establish a network model to study collective behavior of population of neurons. Biological neurons are connected through synapses by transmitting electrical signals, i.e., action potential. A sufficiently strong synaptic input might evoke an action potential in the receiving neuron.

Suppose $u_i(t)$ denotes the membrane potential for the $i$th neuron $(1 \leq i \leq N)$ where $N$ is the network size. The main idea of incorporating synaptic connections between neurons is to modify the variable input current $I_i(t)$ for each individual neuron-$i$ in the population so that it can be represented as a sum of synaptic currents, $I_i^s(t)$, received from the other $N - 1$ neurons, and external stimulus, $I_i^e(t)$. Mathematically,

$$I_i(t) = I_i^s(t) + I_i^e(t) \tag{4}$$

The reader may refer to [9, 13,14] for a comprehensive explanation on how to define $I_i^s(t)$ and $I_i^e(t)$ for IF and HH models.

The simplest form of connectivity is all-to-all (or mean-field) coupling, where each neuron connects equally with every other neuron. In more complex situations such as random coupling, it is useful to introduce a connectivity matrix in the variable $I_i^s(t)$. The matrix entries express whether the pre-synaptic (sending) neuron is connected or not to post-synaptic (receiving) neuron. One of the most common network topologies found in the literature is the Erdős-Rényi model, where each direct connection from pre-synaptic neuron to post-synaptic neuron is randomly chosen based on a specific probability.

Given that the neuron can either send excitatory or inhibitory post-synaptic potential, the neural network considered through this study consists of two sub-networks of excitatory (E) and inhibitory (I) neurons [9], which in this paper is referred to as E-I network. A schematic representation for E-I network is displayed in Fig. 1 [15]. The curve ending in arrow shows direct excitatory connection within E-population, while straight line ending in arrow shows a direct excitatory connection from E to I populations. On the other hand, the curve ending in solid circle shows direct inhibitory connection within I-population, while straight line ending in solid circle shows direct inhibitory connection from I to E populations. Finally, balanced networks are those where the average synaptic input is typically negligible, due to the cancellation between excitation and inhibition.

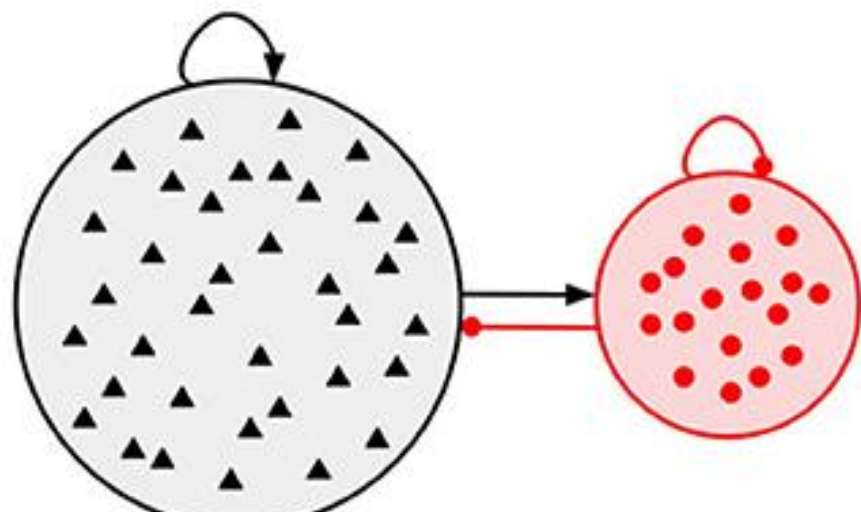

**FIGURE. 1**. *A Schematic Representation for Two Interacting Excitatory (Black Triangles) and Inhibitory (Red Circles) Neuronal Populations*

## METHODOLOGY

A systematic literature review through Preferred Reporting Items for Systematic reviews and Meta-Analyses (PRISMA) is performed to conduct this study [5]. It comprises three main stages, i.e., identification of relevant documents through databases; screening process; and reporting the studies included in the review.

In this study, the SCOPUS database is used as the main source for literature identification due to several reasons. First, SCOPUS includes a vast range of journals across various disciplines, providing a broad and diverse collection of literature. Second, it is generally considered to have an intuitive interface, making it easier for researchers to navigate and find relevant records. The initial search for literatures is conducted by combining the terms "collective AND dynamics AND neurons" (see Fig. 2(a)).

All relevant documents retrieved from SCOPUS are then screened based on the following procedures. At first, we exclude these documents that are not written officially in "English" since it is recognized as a widely used language for international communication. Secondly, we exclude the records other than "Articles" because the research articles typically undergo a peer review process, ensuring that the methodologies and findings meet certain academic standards. Finally, due to the nature of mathematical objects being studied as well as the mathematical properties of dynamical regimes, we then restrict the search to "Mathematics" subject area. All these steps are easy to do through SCOPUS interface refine menu.

**TABLE 1.** Eligibility Criteria for Document Selection based on PICOs

| No | PICOs | Eligibility Criteria |
|---|---|---|
| 1 | Population | E-I neuron networks |
| 2 | Intervention/exposure | Model and network parameters |
| 3 | Comparison | Paradigmatic spiking neuron models [6]:<br>▪ LIF neuron<br>▪ QIF neuron<br>▪ HH neuron |
| 4 | Outcomes | Collective phenomena |

All bibliographic details for each article, including citation information, abstracts, and keywords, are retrieved from the SCOPUS database, and exported in both comma-separated values (CSV) and research information system (RIS) formats. This bibliographic information serves as the foundation for conducting co-occurrence analysis to identify the key topics that arise in the study of spiking neuron dynamics, aligning with the research questions outlined in this paper. Practically, VOSviewer software version 1.6.20 is used to support the analysis of bibliographical data.

Once the screening protocol is completed, we then screen the full text of each article to ensure its content is eligible to be included in the review stage. The eligibility (inclusion) criteria have been set according to the Population, Intervention/exposure, Comparison, Outcomes (PICOs) model as displayed in Table 1.

## FINDINGS AND DISCUSSION

### Selection of Study

As the initial step, combination of the words “collective AND dynamics AND neurons” was typed in the SCOPUS interface (Fig. 2(a)) and we have found n = 491 documents for the interval range (2014, 2024) that match with the given keywords. The number of documents per year is displayed on Fig. 2(b). Though the statistics are fluctuating, the trends show the field has attracted significant interest among researchers in the past ten years.

Fig. 2(c) and Fig. 2(d) display the classification of documents by type and subject area, respectively. Among these records, the article-type documents are the most prevalent, followed by conference paper, review, book chapter, letter, editorial, erratum and notes. The topics of neuron dynamics cover a wide range of subject domains, ranging from mathematics, physics and astronomy, neuroscience, etc. (Fig. 2(d)).

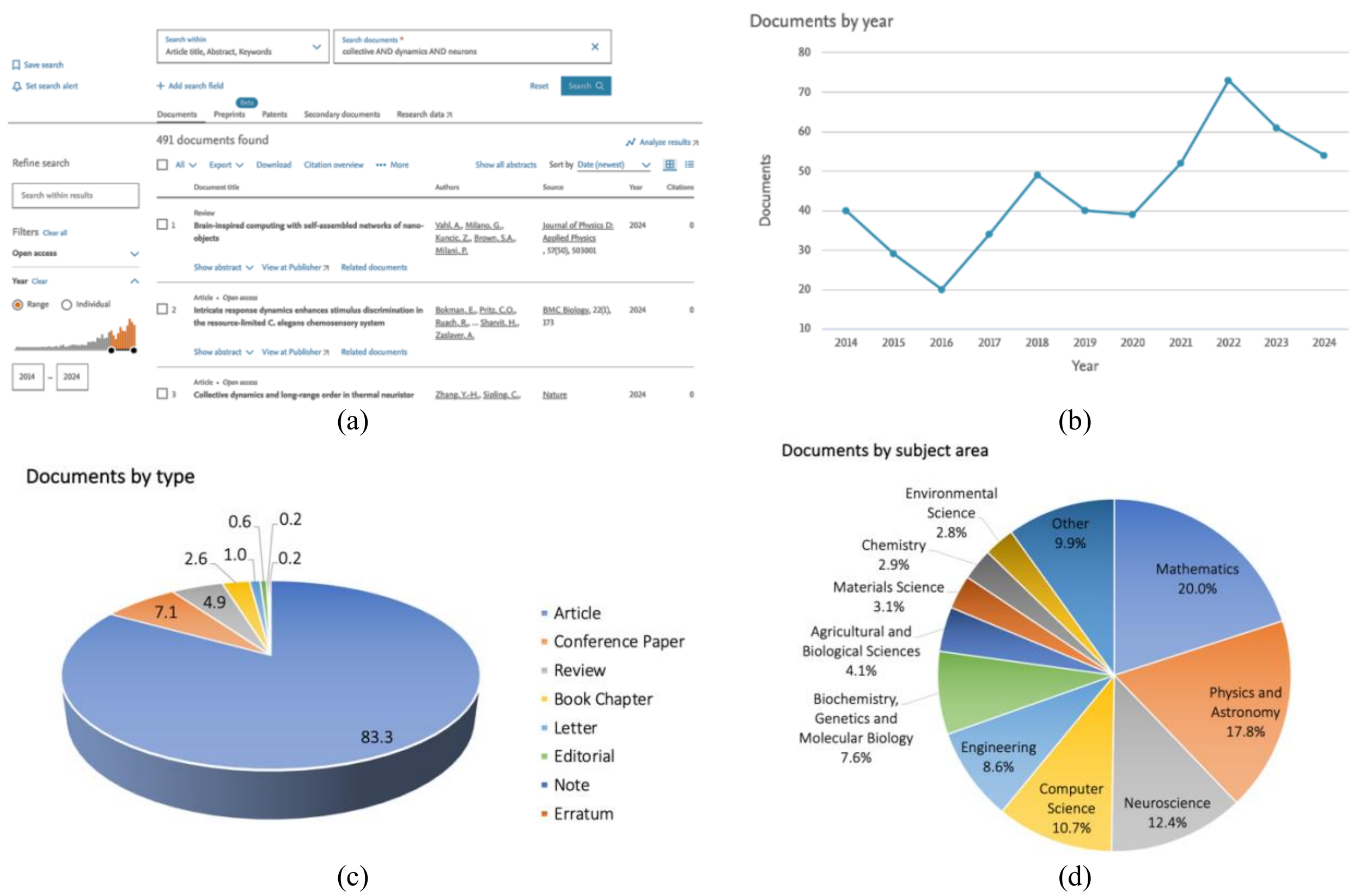

**FIGURE. 2**. Initial Search for the Records 2014-2024: (a) the SCOPUS Interface, (b) Number of Documents, (c) Type of Documents (%), and (d) Documents by Subject Area (%)

The screening process excludes all non-English (n = 10) and non-Article (n = 82) records, resulting in 399 remaining records. The density plot displayed in Fig. 3(a) summarizes the co-occurrence analysis outcomes based on the "Author Keywords" information. The finding highlights the research trends on the dynamics of spiking neural network models and its application. Notice that there are 19 keywords (synchronization, collective dynamics, neural network, bifurcation, memristor, etc.) that occur frequently in the database with "synchronization" is the most frequent keyword used by the authors.

As results of restricting action imposed to "mathematics" subject area, the records reduce to n = 177 documents which are then evaluated for eligibility. The whole process of systematic selection of study is summarized in the Fig. 3(b). Meanwhile, Table 2 presents the distribution of n = 6 eligible documents included in review according to the inclusion criteria in Table 1. Based on the relevant documents, four dynamical regimes have been identified: synchrony, irregular collective dynamics, steady state, and oscillatory behavior. Notably, the QIF neuron model is more frequently utilized in these studies than that of LIF and HH models.

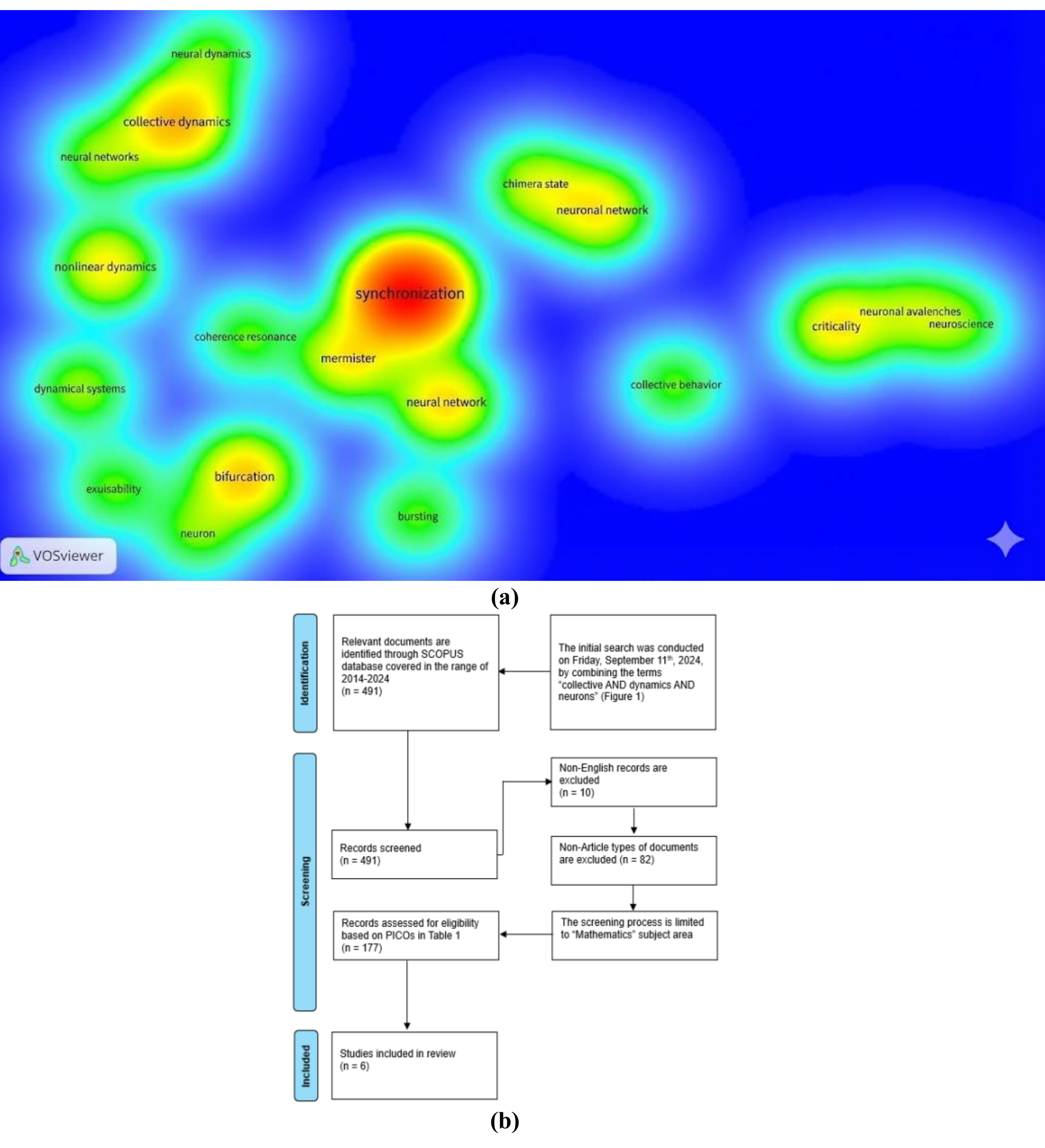

**FIGURE. 3**. (a) Co-occurrence Analysis Outcomes for n = 399 documents, (b) PRISMA Diagram

**TABLE 2.** Distribution of n = 6 Eligible Records

| Dynamical regimes | E-I Neuron Networks | | |
|---|---|---|---|
| | LIF | QIF | HH |
| Synchrony | [16] | [16] | [17] |
| Irregular activity | [18] | [18] | |
| Steady state | | [19,20] | [14] |
| Oscillatory behavior | | [19,20] | [14] |

## Literature Review

In this section we review typical collective behaviors observed in spiking neural network models according to 6 (six) selected studies displayed in Table 2. We address two main points, including (i) the mechanism underlying collective behavior under the setups being studied, and (ii) its possible connection to brain operations. To enrich the discussion, we incorporate several studies recommended by experts.

### *Synchrony*

Synchrony in spiking neural networks refers to the coordinated timing of action potentials (spike) among neurons. This phenomenon is crucial for various brain functions, including perception, learning, and memory. When neurons fire in synchrony, they can enhance signal transmission and processing efficiency. However, excessive synchrony can lead to disorders like epilepsy.

Multiple factors may contribute to the emergence of neuronal synchrony including biophysical properties of individual neuron and network structures. One of the underlying mechanisms, as studied in [16], synchronous pattern underlying gamma-rhythms in fully coupled E-I networks of LIF and QIF neuron driven by external perturbation is achieved due to the cooperative action between internal heterogeneity and phase-response curve. Specifically, the external perturbations that slow down individual neurons give enough time for the inhibition from early spiking neurons to effectively suppress the activity of less excited neurons within the network. Meanwhile, increasing neuronal heterogeneity can advance the synchronization of the whole networks.

Three recognizable synchronization scenarios in a large population of neurons – asynchronized, partial synchronized, and fully synchronized state [7,21]. These states differ in how the temporal fluctuations of a global variable, like population-average voltage (membrane potential) or population-average synaptic conductance, alter with respect to network size $N$. Past study on balance E-I network of LIF neurons [22] has identified typical dynamical states: synchronous regular (SR), asynchronous regular (AR), asynchronous irregular (AI), and synchronous irregular (SI). In the SR regimes, the neurons are in fully synchronized states with regular firing patterns, so they behave like oscillators. In the AR regimes, the global activity is stationary with each individual neuron firing in quasi-regular manner. The AI regimes behave like that of AR with highly irregular individual firing at low rates. In the SI regimes, the global activity behaves as an oscillator with highly irregular individual firing at low rates.

Given that a random network configuration comprises two interacting populations of E-I neurons, synchronization can be observed upon detuning the synaptic weight parameter. As reported in [17] that the population of HH neurons exhibits asynchronous-like patterns when E-subpopulation receives weak positive charges from pre-synaptic cells, while on the other side the inhibitions from I-subpopulation have global effects in shaping the network dynamics. Increasing the synaptic weight (i.e., positive charges are sufficiently strong), will entrain the whole system towards partial synchronization.

### *Irregular activity*

Irregular collective behavior is a dynamical regime of partial synchronization, where a fraction of neurons within the network fire the spike synchronously. In the study reported here [18], irregularity is provoked by balance mechanism: excitatory neurons promote activity by increasing the likelihood of action-potential, while inhibitory neurons reduce activity by suppressing firing. The regime is robust both in LIF and QIF neurons for large and small synaptic weight, strong and weak external currents, given that the network connectivity is random, and each neuron receives a fixed in-degree connection.

In a more generic setup, namely, E-I network of coupled phase oscillators studied in [13], the emergence of collective irregular dynamics is accompanied by several mathematical properties, e.g., (1) The global variables such as average of population phase show a stochastic-like behavior; (2) The power spectra of global variables have a broad shape and it is independent of the network size; and (3) the distribution of phases changes over time, unlike the standard asynchronized state, where the probability distribution remains constant over time.

Balancing excitation and inhibition in E-I network is crucial for maintaining stability and preventing runaway excitation, which can lead to issues like seizures. When excitation and inhibition are well-balanced, the network can process information efficiently, allowing for optimal signal transmission and reducing noise. Disruptions in this balance can result in various neurological disorders. The balance mechanism helps regulate overall network behavior, ensuring that neural circuits function properly.

### *Stationary state*

Stationary (steady) state refers either to an inactive phase of the neuron population that prevents them from firing spikes, or to their inability to fire spikes in a synchronized manner, such as in cases of asynchronous regime. The former scenario follows the general property of an excitable cell: a minimum stimulation charge should be applied to neurons to elicit the spike, otherwise no spike is produced [23]. In the latter one, individual neurons show random firing patterns rather than synchronous firing, which reflects the fact that the sequence of spikes are very weakly correlated and hence the global observable variables, e.g., population firing rate or population-average potential etc. approach time-independent behavior as the network size is getting larger [21].

Weak temporal variations in stimuli across different neurons result in the establishment of a stationary state. For instance, the E-I network of HH neurons studied in [14] indicates that low degrees of average synaptic weight and external stimulation are not sufficient to excite the whole E-population for firing, and that the I-population remains at quiescent state, turning the global dynamics to zero activity. Furthermore, the study of balance E-I network of QIF neurons has reported that the stationary state dynamics persists for distinct types of heterogeneous parameters, including excitable charges, inter and intra-network couplings [19]. In sparse random E-I network of QIF model, three distinct neurons groups have been identified in the asynchronous regime: subthreshold neurons, a group of neurons driven by fluctuation, and mean driven neurons, whose activity is primarily influenced by the in-degree connectivity and mean input currents [24].

The presence of stationary state regimes has advantages in preventing excessive synchronization in the brain, which is often associated with neurological diseases like Perkinson’s disease and epilepsy. A theoretical study has proposed a control strategy on the stimulating charge to achieve a therapeutic effect for the disease [25]. The mathematical frameworks developed therein account for minimizing the side effects to the nerve tissue during the treatment procedure. The study is extended to the E-I network of heterogeneous QIF neurons [20] which indicates that synchronous dynamics can be suppressed either by perturbing I-population with high-frequency stimulation or applying a single inhibitory finite-width pulse to E-population. When a high-frequency stimulation is applied, the proportion of neurons in I-population increases due to the variation of neurons intrinsic properties, which then disturbs synchronous firing activity of the whole network and hence stabilize the resting state.

### *Oscillatory dynamics*

Oscillation is a prevalent phenomenon in neuronal systems. The oscillatory patterns observed in EEG recordings can occur at different frequencies, categorized into bands: delta (1.5–4 Hz), theta (4–8 Hz), alpha (8–12 Hz), beta (12–30 Hz), and gamma (greater than 30 Hz) [26, 27]. These rhythms correspond to electrical activities in the brain. While the origins of these rhythms remain debatable, there are many signs of progress in computational studies to uncover the underlying neural mechanisms [28]. In single neuron model, oscillation is achieved when the constant input charge, characterized by variable $I(0) = I_0$ in Eqns. (1), (2), and (3), exceeds a critical value [13]. In such cases, the neuron fires a spike regularly in time, leading to the consistent pattern of inter-spike intervals.

In a network level, the oscillatory patterns are believed to arise from synchronized firing activity amongst neurons, according to experimental study [29]. The utilization of mean-field equations, formulated in terms of a few macroscopic variables derived from spiking neural networks, facilitates a more tractable approach for implementing bifurcation theory to investigate the emergent of collective oscillations (COs). It has been reported that the COs emerge via Hopf bifurcation upon varying the width of synaptic pulses in single population of QIF neurons [30]. The generation of COs, including gamma rhythms, within a purely inhibitory QIF neuronal population is controlled by finite duration of inhibitory post-synaptic potentials [31, 32]. A brief instantaneous stimulation can switch the gamma oscillations between slow and fast states [33]. The fast gamma rhythm arises from coordinated tonic neural firing, while the slow one is driven by intrinsic fluctuations resulting from irregular neural activity. In E-I networks of QIF model, the ratio of excitatory and inhibitory synaptic time scales can support weakly modulated gamma oscillation or strongly synchronized gamma oscillation [34]. Moreover, the influence of external drive in facilitating collective oscillations (COs) within spiking neurons is of significant importance. Modulating the intensity of external currents in excitatory-inhibitory (E-I) networks, encompassing both QIF and HH neurons, induces various forms of COs, including simple periodic oscillations, aperiodic dynamics, and chaotic states [14, 24]. These regimes are robust against heterogeneous intra, and inter-network parameter couplings [19].

The simultaneous presence of different frequency bands of rhythmic oscillations often leads to cross-frequency couplings (CFCs) phenomena, in which a high frequency oscillatory dynamic is stimulated by lower frequency band. The common example is gamma modulation by theta rhythm, and its variance includes phase-phase and phase-amplitude CFC, as displayed in two inhibitory QIF populations connected via bidirectional coupling [31]. The same scenario of CFC also exists in two neuronal modules (each module comprises E-I network of QIF neurons) coupled with a delayed conduction-based synapse [35]. Studies have shown that in both ING (Interneuron Network Gamma) and PING (Pyramidal-Interneuron Network Gamma) systems influenced by sinusoidal input, CFC can produce two distinct forms of nested oscillations. One form exhibits perfectly locked states between theta and gamma rhythms, leading to periodic behaviour. The other form displays imperfectly locked states, resulting in quasi-periodic and even chaotic patterns [36].

The inclusion of synaptic-based Working Memory (WM) and spike-frequency adaptation (SFA) in QIF neural networks enrich the emerging dynamics. In the first setup [37], adjusting the frequency of an external periodic drive for single excitatory QIF neurons, specifically within the delta and beta ranges, influences transitions between low and high activity states. The high activity state, associated with "recall or working memory," arises from delta band (slow) periodic stimulation of the network. Conversely, the low activity state, representing "clearance or ground state," results from beta band (high) periodic stimulation. The high activity state in E-I networks, known as "memory loading," is characterized by population bursting activity in beta-gamma band. It is triggered by selective reactivation of excitatory populations and then sustained by inhibitory neurons [38]. In the second setup [39], single population of QIF neurons with SFA exhibits population bursting in the excitatory networks and tonic spiking in the inhibitory ones. Without SFA, symmetric coupling in E-E populations results in anti-phase spiking activity between the two populations, while in I-I populations, it gives rise to symmetric and asymmetric stationary state of population firing rate, as well as chimera-like solutions. The addition of SFA in both I-I and E-E populations gives rise to novel macroscopic dynamics, encompassing symmetric and asymmetric stationary states, nested oscillatory patterns, coexistence regimes of COs, bursting phenomena, and theta-gamma CFCs.

## CONCLUSION AND LIMITATION

Over the last ten years, there has been a growing trend in the study of collective dynamics in spiking neural networks. The growing recognition of the importance of the study in understanding brain function and behavior, combined with technological advancements (such as machine learning and artificial intelligence) and interdisciplinary research collaboration, is driving interest in this research theme. Through the PRISMA procedure and bibliometric analysis, we chose six pertinent research articles on the theme that meet the criteria for review based on the research questions outlined in this study. Four robust dynamical regimes have been identified: synchrony, irregular behavior, stationary state, and oscillatory dynamics. Our review outcomes indicate that the collective dynamics in E-I spiking neurons arise from the interplay of intrinsic neuronal properties, network topology, and the nature of external stimuli.

The widespread use of QIF neuron models in the literature highlights their important role, particularly in understanding collective behavior of neurons. The QIF model offers mathematical tractability and lower computational cost compared to the biophysical HH model. In larger network sizes, it is amenable to derive mean-field equations of QIF model and study the emerging dynamics through bifurcation analysis. The analysis is far more challenging to do with HH model due to its complexity of ion-channel mechanisms. While the QIF model is more computationally expensive than the LIF model, it can capture nonlinear behaviors near the threshold for spike generation due to its quadratic term. This allows it to represent richer dynamics compared to the linear membrane potential changes in the LIF model.

In conclusion, we wish to emphasize the limitations of this study. We relied solely on the SCOPUS database for our initial search, which resulted in a limited number of relevant articles. In the future, it will be important to utilize additional databases, such as Web of Science and Science Direct, to expand the literature review. On top of that, refining the eligibility criteria for literature search is the most important point since it determines the relevant documents to be included in the review stage. Finally, employing automation tools (like CADIMA, Rayyan, etc.) for the screening process is strongly recommended to enhance efficiency and ensure objective results.

## ACKNOWLEDGMENTS

The authors would like to thank Simona Olmi for the meaningful suggestion to the paper. The authors acknowledge support from the Fundamental Research Grant Scheme with Project Code: FRGS/1/2022/STG06/USM/02/1 by the Ministry of Higher Education, Malaysia (MOHE).